\documentclass[aps,prb,reprint,groupedaddress,floatfix]{revtex4-2}

\usepackage{graphicx}
\usepackage{amsmath,amssymb}
\usepackage{dcolumn}
\usepackage{mathrsfs}
\usepackage{physics}
\usepackage{booktabs}
\usepackage{color}
\usepackage{xurl}
\usepackage{hyperref}

\begin{document}



\title{The Dielectric Bowtie Effect: Classical Electromagnetic Edge Singularities in Subwavelength Cavities}



\author{Valdemar Bille-Lauridsen}
\email[]{vchbi@dtu.dk}
\affiliation{Department of Electrical and Photonics Engineering, Technical University of Denmark, Building 343, 2800 Kongens Lyngby, Denmark}
\affiliation{NanoPhoton - Center for Nanophotonics, Technical University of Denmark, Building 343, 2800 Kongens Lyngby, Denmark}

\author{Jesper M\o rk}
\affiliation{Department of Electrical and Photonics Engineering, Technical University of Denmark, Building 343, 2800 Kongens Lyngby, Denmark}
\affiliation{NanoPhoton - Center for Nanophotonics, Technical University of Denmark, Building 343, 2800 Kongens Lyngby, Denmark}


\date{\today}

\begin{abstract}
Dielectric bowtie nanocavities can concentrate light into subwavelength regions without the ohmic losses of plasmonic metals. We show that this enhancement is the finite-geometry realization of a classical electromagnetic edge singularity. Unlike an isolated dielectric wedge, the scaling in a bowtie is governed by an exponent determined by a collective four-sector singularity. 
In a finite structure, this scale-free singular field is regularized by the gap size, while the bowtie length sets the outer scale. The tip radius, gap, and bowtie length therefore play distinct physical roles: curvature cuts off the local wedge singularity, the gap cuts off the collective bowtie singularity, and the outer length sets the range over which the field can build up. Electrostatic simulations confirm the predicted scaling laws, while three-dimensional quasinormal-mode simulations show how the same near-field mechanism is accessed and limited by realistic dielectric nanocavities. 

\end{abstract}


\maketitle



A dielectric optical nanocavity enhances light-matter interaction through the
Purcell effect by confining electromagnetic energy in a resonator with high
quality factor $Q$ and small mode volume $V$ \cite{Purcell_PR_1946}. Photonic-crystal
(PhC) defect cavities confine light through a photonic bandgap and are limited
to mode volumes above the diffraction limit, $V > V_\lambda = (\lambda/2n)^3$
\cite{akahaneHighQPhotonicNanocavity2003,deotareHighQualityFactor2009}. This was
long regarded as a fundamental limit for dielectric nanocavities, surpassed only
by plasmonic metal cavities \cite{khurgin2015loss,gramotnev2010plasmonics,krasnok2022low}. Theoretical designs
\cite{robinsonUltrasmallModeVolumes2005,huDesignPhotonicCrystal2016,choiSelfSimilarNanocavityDesign2017,wang2018maximizing}
and, more recently, experimental investigations
\cite{huExperimentalRealizationDeep2018,albrechtsenNanometerscalePhotonConfinement2022,babar2023self,xiongExperimentalRealizationDeep2024,ouyangSingularDielectricNanolaser2024}
have reported mode volumes well below $V_\lambda$. Such cavities all
rely on deep-subwavelength dielectric structuring and define a new class of Extreme Dielectric Confinement (EDC) cavities. When optimized for minimum mode volume or, equivalently,
a large local density of optical states, a bowtie shape appears to be the generic geometry that emerges
\cite{choiSelfSimilarNanocavityDesign2017,wang2018maximizing,mignuzzi2019nanoscale,christiansenOrdersMagnitudeReduction2026}. 

\begin{figure}[ht]
    \centering
    \includegraphics[width=0.9\linewidth]{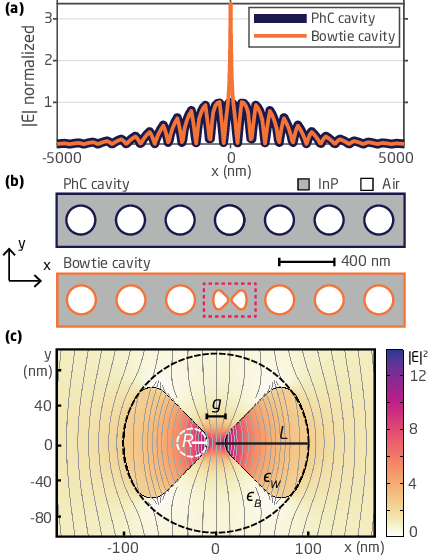}
    \caption{(a) Electric field in a PhC nanobeam cavity and in one where a bowtie
replaces the central hole, normalized to the PhC maximum. (b) Schematic of the
two geometries. (c) Finite bowtie geometry studied here, set by the tip radius
$R$, gap $g$, half-opening angle $\alpha$, and length $L$ (here $g=30$~nm,
$R=20$~nm, $L=100$~nm, $\alpha=45^\circ$).}
    \label{fig:Motivation}
\end{figure}

Figure~\ref{fig:Motivation}
shows an illustrative example: replacing the central hole of a nanobeam PhC cavity \cite{quanDeterministicDesignWavelength2011} with a
bowtie raises the peak field intensity by an order of magnitude while leaving
the surrounding mode almost unchanged \cite{halimiControllingModeProfile2020}. Because the mechanism is purely
dielectric, it avoids the ohmic losses of plasmonic resonators
\cite{kwonDesignPlasmonicCavities2014} and provides radiative emission enhancement \cite{cordova2024single} to the benefit of a range of optical
devices, including nanolasers \cite{xiongNanolaserExtremeDielectric2025},
switches \cite{dong2026enhancement}, and single-photon sources
\cite{choiSelfSimilarNanocavityDesign2017}. Yet a unified account of why they
concentrate light so strongly is still missing. The physical pictures proposed
so far each capture only part of the mechanism: slot and anti-slot boundary
conditions at refractive-index discontinuities
\cite{robinsonUltrasmallModeVolumes2005,huDesignPhotonicCrystal2016,choiSelfSimilarNanocavityDesign2017}, modes with
large imaginary wave vectors \cite{ouyangSingularDielectricNanolaser2024}, and
dielectric lightning-rod effects \cite{albrechtsenTwoRegimesConfinement2022}.
None of them identifies the bowtie field as the finite realization of a
classical dielectric edge singularity, a problem studied extensively in
electrostatics
\cite{meixner1972,davisElectostaticEdgeModes1976,marxComputedFieldsEdge1990,idemenConfluentTipSingularity2003}.

\begin{figure}[t]
\centering
\includegraphics[width=0.9\linewidth]{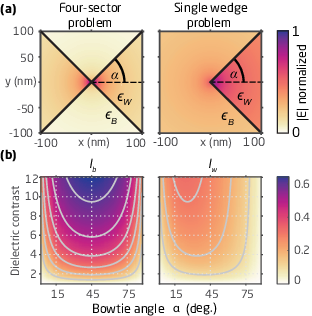}%
\caption{(a) Field and geometry of the ideal four-sector and single-wedge problems for dielectric contrast $\epsilon_{\mathrm{B}}/\epsilon_{\mathrm{W}}=9$ and opening angle $\alpha=45^\circ$. (b) Singularity exponents $l_{\mathrm{w}}$ (single wedge) and $l_{\mathrm{b}}$ (four sector) as a function of $\alpha$ and $\epsilon_{\mathrm{B}}/\epsilon_{\mathrm{W}}$.
}
\label{fig:Geometries}
\end{figure}

Here, we show that the field enhancement at the center of a bowtie is set not by a local single-wedge or
lightning-rod singularity but by a collective four-sector singularity, whose
exponent $l_{\mathrm{b}}$ is fixed by the half-opening angle and the dielectric
contrast and generally differs from the single-wedge exponent $l_{\mathrm{w}}$.
In a finite bowtie [Fig.~\ref{fig:Motivation}(c)], the scale-free field of the
ideal four-sector problem [Fig.~\ref{fig:Geometries}] is regularized by three
distinct lengths, each with a separate role: the tip radius $R$ cuts off the
local wedge singularity, the gap $g$ cuts off the collective four-sector
singularity, and the bowtie length $L$ sets the outer scale over which the
singular field builds up. The center-field intensity then obeys
\begin{equation}
    |E(0,0)|^2 \propto \left(\frac{L}{g}\right)^{2l_{\mathrm{b}}},
\label{eq:center_field_scaling}
\end{equation}
a scaling we confirm with electrostatic simulations and three-dimensional
quasinormal-mode calculations. The bowtie effect is thus the regularization of a
classical singularity rather than a new singular regime
\cite{maoSingulonicsNarwhalshapedWavefunctions2025}, and the outer length $L$, largely overlooked in earlier work, proves as essential to the enhancement as the gap $g$. This picture also clarifies how the diffraction-limited field of a conventional wavelength-scale cavity combines with the near field of the bowtie singularity to set the final mode profile that governs the light-matter
interaction.

Before studying a realistic bowtie geometry, we first recall the scale-free electrostatic singularities that determine the possible power-law field scalings \cite{ouyangSingularDielectricNanolaser2024}. In the quasistatic limit, the potential satisfies Laplace's equation, and sharp dielectric corners admit separable power-law solutions. The single dielectric wedge problem, Fig.~\ref{fig:Geometries}, has been solved in this form in the classical literature on electromagnetic edge singularities \cite{meixner1972,davisElectostaticEdgeModes1976,boardmanElectrostaticEdgeModes1985,idemenConfluentTipSingularity2003}. It gives a local lightning-rod enhancement to the field near an isolated tip,
\begin{equation}
    |E(r)| \propto r^{-l},
    \label{eq:powerlaw}
\end{equation}
where $r$ is the radial distance from the tip, and the exponent $l=l_{\mathrm{w}}$ depends only on the opening angle and dielectric contrast.

The touching-bowtie limit can be treated in the same way, but with the angular domain replaced by a four-sector dielectric geometry. This geometry is still scale free and therefore gives an analogous collective power law, with $l_{\mathrm{w}}$ replaced by a bowtie exponent $l=l_{\mathrm{b}}$ determined by a four-sector eigenvalue problem, see the End Matter. The important point is that $l_{\mathrm{b}}$ is generally different from $l_{\mathrm{w}}$, because the two opposing tips are coupled through the full bowtie geometry rather than acting as two independent wedges. 

Figure~\ref{fig:Geometries} compares the two exponents and the corresponding fields. 
Both exponents increase with the dielectric contrast and depend on the opening angle. The four-sector exponent $l_{\mathrm{b}}$ is maximal at a full opening angle $2\alpha=90^\circ$, coinciding with the maximal center enhancement found in earlier numerical studies \cite{albrechtsenTwoRegimesConfinement2022}. This already suggests that the bowtie enhancement is governed by $l_{\mathrm{b}}$ rather than by the single-wedge exponent $l_{\mathrm{w}}$.

\begin{figure}
    \centering
    \includegraphics[width=0.95\linewidth]{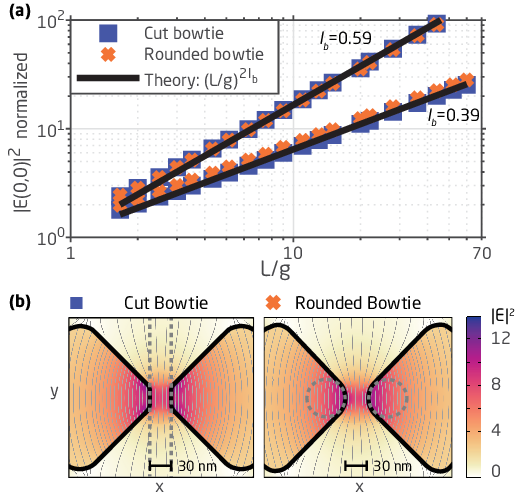}
    \caption{
    (a) Electric field intensity at the bowtie center for different geometries, where the bowtie length $L$ and the gap size $g$ are varied independently. The slopes of the solid lines are given by the four-sector singular exponents determined by the two geometries;  $l_{\mathrm{b}}=0.59$ for $\alpha=45^\circ$ and $\epsilon_{\mathrm{B}}/\epsilon_{\mathrm{W}}=3$, and $l_{\mathrm{b}}=0.39$ for $\alpha=60^\circ$ and $\epsilon_{\mathrm{B}}/\epsilon_{\mathrm{W}}=2$. For the rounded bowtie, the radius is set to $R=g$.
    (b) The electric field inside the bowtie gaps for a geometry with locally round bowtie tips (Rounded Bowtie) and a geometry with locally flat bowtie tips (Cut Bowtie). 
    }
    \label{fig:centerfield_scaling}
\end{figure}

The ideal singularity fixes the exponent but not the amplitude, since the scale-free four-sector solution acquires a definite amplitude only when matched to a finite geometry. To make this matching concrete, we perform two-dimensional electrostatic simulations of a finite dielectric bowtie embedded in a large uniform background of relative permittivity $\epsilon_{\mathrm{B}}$ using COMSOL Multiphysics \cite{COMSOL63}. 
A fixed potential difference applied between the top and bottom boundaries generates a uniform field of 1~V/m far from the bowtie, which thereby sets the normalization for the center-field intensity enhancement, $|E(0,0)|^2$, caused by the bowtie. This is the quantity of interest, since it determines the coupling to a dipole emitter placed in the gap. Figure~\ref{fig:centerfield_scaling} shows that, when plotted as a function of the ratio $L/g$, the results for different geometries fall onto the scaling predicted by Eq.~\eqref{eq:center_field_scaling}, with an exponent set by $2l_{\mathrm{b}}$, as determined from the four-sector eigenvalue problem. Here, $L$ is varied between 100 nm and 300 nm, and $g$ is independently varied between 5 nm and 60 nm.

\begin{figure}
    \centering
    \includegraphics[width=0.98\linewidth]{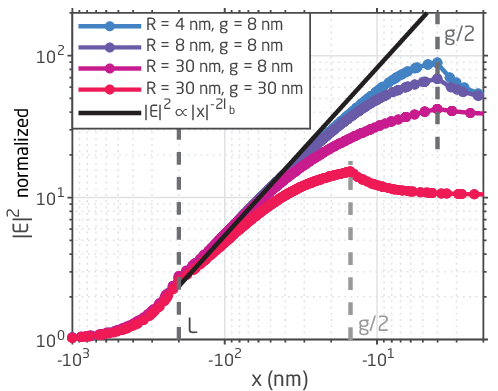}
    \caption{Center-line field profiles of finite bowtie structures obtained from two-dimensional electrostatic simulations. The black curve shows the scaling predicted by the ideal four-sector problem. The colored curves show four finite geometries. The parameters are $\alpha=45^\circ$ and $\epsilon_{\mathrm{B}}/\epsilon_{\mathrm{W}}=3$. For the first three geometries, only the radius $R$ is varied.
    }
    \label{fig:Regions_field_growth}
\end{figure}

The origin of this single-variable scaling is seen directly in the center-line profiles of Fig.~\ref{fig:Regions_field_growth}, which separate into three regions. Outside the bowtie, $|x|\gtrsim L$, the field is set by the background and does not follow the singular law. In the bowtie region, $g/2\lesssim|x|\lesssim L$, the field grows toward the gap as $|E|^2\sim|x|^{-2l_{\mathrm{b}}}$, the power law predicted by the four-sector problem. Inside the gap, $|x|<g/2$, the two tips are resolved and the field increases close to the tips but saturates smoothly in the center, between the rounded interfaces. The enhancement is therefore accumulated in the bowtie region, not in the gap itself: the field grows from the outer scale $L$ inward according to the four-sector law and is cut off when it reaches the gap $g$. This is precisely what the ratio $L/g$ measures, and it fixes the physical roles of the two lengths. Increasing $L$ lengthens the range over which the field can grow, while decreasing $g$ moves the inner cutoff closer to the ideal touching-bowtie singularity.

It follows that there is no universal gap law of the form $|E(0,0)|^2\propto g^{-1}$. Reducing $g$ raises the center field, but the exponent of that increase is set by the collective singularity. The same reasoning rules out a universal contrast law $|E(0,0)|^2\propto m^q$ with constant $q$, where $m=\epsilon_{\mathrm{B}}/\epsilon_{\mathrm{W}}$. For $\alpha=45^\circ$ the model gives $q(m)=1+[4\sqrt{m}/\pi(m+1)]\log(L/g)$, so the apparent contrast exponent depends on the cutoff ratio and saturates at large $m$, giving a natural interpretation of the scaling seen in earlier work \cite{christiansenOrdersMagnitudeReduction2026}.

Figure~\ref{fig:centerfield_scaling} also distinguishes the collective enhancement from a purely local lightning-rod effect. As seen, a geometry that is locally flat (referred to as "cut bowtie") rather than rounded in the gap gives the same center field scaling, so the bowtie effect is not due to a sharp tip or related to the single-wedge problem and exponent $l_{\mathrm{w}}$. We notice that the locally flat designs also emerge from topology optimization for extended emitters \cite{bille-lauridsenDeterministic2026}, rather than the dipole emitters targeted by mode volume minimization. The tip radius $R$ plays the separate role of regularizing the local single-wedge singularity at each interface, for which the exact rounded-tip solution is known \cite{boardmanElectrostaticEdgeModes1985}. It strongly affects the peak field at the tip but is secondary for the center field provided $R\lesssim g$, as the four cases in Fig.~\ref{fig:Regions_field_growth} show. In the hierarchy $R\lesssim g\ll L$, the single wedge problem is only related to the local lightning-rod field, while the center field is dominated by the collective enhancement built up over $g<r<L$. 

\begin{figure*}
    \centering
    \includegraphics[width=0.9\linewidth]{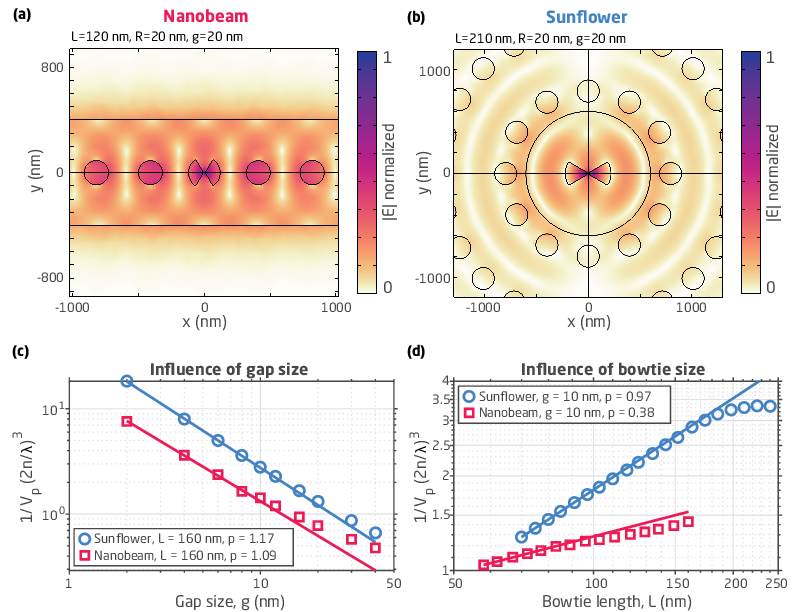}
    \caption{
    (a) The QNM field distribution around a bowtie in a nanobeam cavity. (b) The corresponding QNM field distribution in a bowtie sunflower cavity.
    (c) Inverse mode volume from 3D QNM simulations of nanobeam and sunflower cavities in units of $(2n/\lambda)^3$ as a function of gap size $g$, with $R=g$. The lines show power-law fits $1/V_p \propto (L/g)^p$ to the smallest-gap points. The four-sector problem predicts $p=2l_b=1.18$.
    (d) Similarly, the inverse mode volume is plotted and fitted as a function of bowtie size $L$.
    }
    \label{fig:E00_scaling_3dQNM}
\end{figure*}

The electrostatic theory above gives the singular bowtie response when driven by a uniform electric field. In a nanocavity, this field may not be uniform, as it is supplied by a three-dimensional quasinormal mode (QNM), which provides the resonance, fixes the global normalization, and sets the field incident on the bowtie. We therefore compare the singular scaling to the inverse QNM mode volume
\begin{equation}
    \frac{1}{V_p}
    =
    \operatorname{Re}\left(\frac{\epsilon |E(0,0)|^2}{\langle\!\langle E|E\rangle\!\rangle}\right),
\end{equation}
where $\langle\!\langle E|E\rangle\!\rangle$ is the QNM normalization \cite{kristensenGeneralizedEffectiveMode2012}. This quantity is the 3D cavity analogue of the electrostatic center-field intensity, since it measures the field at the emitter position relative to the normalized mode energy.

Fig.~\ref{fig:E00_scaling_3dQNM} shows examples of full three-dimensional simulations of the field in two types of EDC cavities, i.e., the nanobeam geometry studied so far and a so-called sunflower cavity \cite{Ma:24}. 

The gap sweep in Fig.~\ref{fig:E00_scaling_3dQNM}(c) gives the most direct test of the singular cutoff. At fixed bowtie size, reducing $g$ changes the inner regularization of the four-sector singularity while leaving the outer length and the QNM envelope nearly unchanged. The inverse mode volume therefore follows a power-law trend close to the electrostatic prediction $1/V_p \propto\left(L/g\right)^p$ for $p \sim 2l_{\mathrm{b}}=1.18$
over the smallest gaps. For larger $g$, the hierarchy between the gap and other characteristic length scales of the system is reduced, and the apparent exponent decreases.

The $L$ sweep in Fig.~\ref{fig:E00_scaling_3dQNM}(d) examines a different part of the problem. Changing $L$ changes the outer scale of the singular buildup and also changes how the bowtie samples the spatial QNM envelope. The electrostatic scaling with $L$ is therefore recovered only while the cavity field is nearly uniform over the bowtie. More generally, the relevant outer scale is an effective length $L_{\mathrm{eff}}$, set by both the physical bowtie size and by the QNM field variation length $L_{\mathrm{QNM}}$. The effective length will approach the geometric length $L$ for a slowly varying QNM, but grows more slowly once the bowtie extends into regions of stronger field gradients. The QNM therefore sets the outer matching condition for the finite singularity.

This interpretation explains the difference between the two cavities. The sunflower cavity supports a broader QNM near the center, so increasing $L$ extends the region over which the bowtie collects and concentrates the field, and the scaling remains closer to the electrostatic result. The nanobeam QNM is more compact, so the singular buildup saturates earlier as the bowtie extends into a region of steeper field gradient.

The saturation of the sunflower inverse mode volume around $L\sim200\,\mathrm{nm}$ gives a direct example of this matching. From Fig.~\ref{fig:E00_scaling_3dQNM}(b), the bowtie edge lies close to a QNM field node for $L=210\,\mathrm{nm}$, so further increasing $L$ adds little drive to the central singularity.

The apparent exponent also depends on the full set of parameters. Reducing $g$ and $L$ in the nanobeam shifts the mode volume scaling towards a stronger $L$ dependence, with an apparent exponent of order $p\sim0.7$ for $L=40\,\mathrm{nm}$ and $g=R=1\,\mathrm{nm}$. The observed scaling is therefore co-determined by $g$, $L$, $R$, and the QNM envelope. The singular problem only fixes the local power law, which is accessed most clearly when the hierarchy $R\lesssim g\ll L < L_{\mathrm{QNM}}$ is well satisfied.

These results establish a separation of scales in the singular description: the electrostatic problem fixes the local exponent $l_{\mathrm{b}}$ and thereby the center-field enhancement, while the QNM normalization and the spatial matching between the bowtie and the host cavity determine how much of this enhancement is realized in the mode volume. The problem is thus separable only at the level of the local exponent; the finite mode volume is a property of the combined bowtie-QNM system.

This separability translates directly into a design rule for a centered emitter: the gap should be minimized, the rounding radius kept small enough that it does not replace the gap as the dominant cutoff, and the bowtie enlarged over the region where the QNM sustains a high field. The optimal enhancement in a given cavity is thus obtained by matching the bowtie size to the spatial structure of the resonant mode.

The QNM simulations show how the electrostatic singularity enters a realistic nanocavity. The bowtie field is an evanescent near-field enhancement enabled by dielectric edge singularities \cite{meixner1972,marxComputedFieldsEdge1990,budaevElectromagneticFieldSingularities2007,idemenConfluentTipSingularity2003}, which clarifies the relation to diffraction-limit arguments \cite{krasnok2022low}, and the same singular physics links material choice, bowtie angle, gap, radius, bowtie size, and QNM envelope to the achievable device mode volume. Extreme dielectric confinement arises when a global resonant mode delivers field to a local singular geometry that concentrates it below the scale of the wavelength.

\begin{acknowledgments}
This work was supported by the Danish National Research Foundation through NanoPhoton - Center for Nanophotonics, Grant No. DNRF147.
\end{acknowledgments}


\bibliography{references_abbreviated_journals}

\section*{End Matter}

\subsection{Four-sector singularity}
\label{sec:4sector_theory}

The single-wedge solution describes the field close to an isolated tip. On scales $r\gg g,R$, however, the two tips are no longer resolved individually. The bowtie then approaches the ideal touching-bowtie geometry shown in Fig.~\ref{fig:Geometries}(a). This scale-free problem is defined by two dielectric sectors of angular width $2\alpha$ and permittivity $\epsilon_{\mathrm{W}}$, separated by two background sectors of angular width $\pi-2\alpha$ and permittivity $\epsilon_{\mathrm{B}}$.

As for the wedge case, the electrostatic potential is written in separable form,
\begin{equation}
    \Phi(r,\theta)=r^{\lambda}f(\theta).
\end{equation}
Substitution into Laplace's equation gives the angular eigenvalue equation
\begin{equation}
    \frac{d^2 f}{d\theta^2}+\lambda^2 f=0,
\end{equation}
so that within each sector $j$,
\begin{equation}
    f_j(\theta)=A_j\cos(\lambda\theta)+B_j\sin(\lambda\theta).
\end{equation}
The constants in neighboring sectors are related by continuity of the potential and of the normal displacement field at each angular interface,
\begin{equation}
    f_j=f_{j+1},
    \qquad
    \epsilon_j\frac{d f_j}{d\theta}
    =
    \epsilon_{j+1}\frac{d f_{j+1}}{d\theta}.
\end{equation}

It is convenient to collect these two continuous quantities in the state vector
\begin{equation}
    \mathbf{x}(\theta)=
    \begin{pmatrix}
        f(\theta) \\
        \epsilon f'(\theta)
    \end{pmatrix}.
\end{equation}
Across a sector of permittivity $\epsilon$ and angular width $\Theta$, this state evolves as
\begin{equation}
    \mathbf{x}(\theta+\Theta)=P(\lambda,\epsilon,\Theta)\mathbf{x}(\theta),
\end{equation}
with
\begin{equation}
    P(\lambda,\epsilon,\Theta)=
    \begin{pmatrix}
        \cos(\lambda \Theta) & \dfrac{\sin(\lambda \Theta)}{\epsilon\lambda} \\
        -\epsilon\lambda\sin(\lambda \Theta) & \cos(\lambda \Theta)
    \end{pmatrix}.
\end{equation}
One full revolution around the four-sector geometry is therefore described by
\begin{equation}
\begin{aligned}
    T(\lambda)=
    &P(\lambda,\epsilon_{\mathrm{W}},2\alpha)
    P(\lambda,\epsilon_{\mathrm{B}},\pi-2\alpha) \\
    &\times
    P(\lambda,\epsilon_{\mathrm{W}},2\alpha)
    P(\lambda,\epsilon_{\mathrm{B}},\pi-2\alpha).
\end{aligned}
\end{equation}
Periodic continuation around the origin requires $\mathbf{x}(2\pi)=\mathbf{x}(0)$, and hence a nontrivial solution exists only when
\begin{equation}
    \det\!\left[T(\lambda)-I\right]=0 .
    \label{eq:bowtie_eigenvalue}
\end{equation}
Equivalently, since each propagator is unimodular, the condition may be written as $\operatorname{Tr}T(\lambda)=2$. This transcendental equation together with the condition $0<\lambda<1$ determines the allowed four-sector exponent. The electric field scales as $|\mathbf{E}|\sim r^{\lambda-1}$, so the bowtie exponent is
\begin{equation}
    l_{\mathrm{b}}=1-\lambda .
\end{equation}
The resulting $l_{\mathrm{b}}$ is shown in Fig.~\ref{fig:Geometries} together with the single-wedge exponent. Because $l_{\mathrm{b}}$ follows from the full angular boundary-value problem, it is generally different from $l_{\mathrm{w}}$. The bowtie is therefore not simply two independent wedge singularities, but a distinct four-sector singularity. A similar approach has previously been used to study a related plasmonic geometry, including the effect of rounded corners \cite{Jin:13}.

\end{document}